\documentclass{article}
\usepackage{times}
\usepackage{bm}
\usepackage{float}
\usepackage{amsfonts}
\usepackage{graphicx}
\usepackage{amssymb}
\usepackage[T1]{fontenc}
\usepackage{color}

\begin{document}
\title{Crystal Growth as an Excitable Medium}

\author{Julyan H. E. Cartwright$^1$, Antonio G. Checa$^2$, \\ Bruno Escribano$^3$, C. Ignacio Sainz-D{\'\i}az$^1$ \\ \\
$^1$Instituto Andaluz de Ciencias de la Tierra, \\CSIC--Universidad de Granada, \\Campus Fuentenueva, E-18071 Granada, Spain \\
$^2$Departamento de Estratigraf\'{\i}a y Paleontolog\'{\i}a, \\Facultad de Ciencias, Universidad de Granada, \\E-18071 Granada, Spain \\
$^3$Basque Center for Applied Mathematics, \\Alameda de Mazarredo 14, E-48009 Bilbao, \\Basque-Country, Spain
}

\date{version of \today}

\maketitle

\abstract{
Crystal growth has been widely studied for many years, and, since the pioneering work of Burton, Cabrera, and Frank, spirals and target patterns on the crystal surface have been understood as forms of tangential crystal growth mediated by defects and by two-dimensional nucleation. Similar spirals and target patterns are ubiquitous in physical systems describable as excitable media. Here we demonstrate that this is not merely a superficial resemblance; that the physics of crystal growth can be set within the framework of an excitable medium, and that appreciating this correspondence may prove useful to both fields. Apart from solid crystals, we discuss how our model applies to the biomaterial nacre, formed by layer growth of a biological liquid crystal.}


\section{Introduction}
\label{}

Molecular-scale spiral and target patterns on the surface of crystals were predicted even before they were seen directly with microscopes. Burton, Cabrera, and Frank proposed a dynamical model to explain how crystals should grow layer by layer through the addition of material at kinks; growth sites at the edges of a spiral or target pattern \cite{Burton:1951p2792}. The Burton--Cabrera--Frank (BCF) model has since been amply confirmed by observations, although today the quantitative dynamics of crystal growth is known to be affected by a host of effects not considered in the minimal BCF model.
On the other hand, spiral and target patterns similar to those on the surface of crystals are seen in chemistry, e.g., in the Belousov--Zhabotinsky reaction; in biology, e.g., in yeast growth or in heart fibrillation; and in many other instances. 
The latter phenomena are all examples of excitable media, a rather general class of systems in which elements are quiescent until excited by some stimulus, after which they are unresponsive to further stimuli during some refractory period before returning to their initial quiescent, excitable state \cite{Ball:2001p3015,meron}. This very simple physics is sufficient to produce complex spatiotemporal patterns of targets and spirals, which is why these patterns are observed so generally in many different fields \cite{Tyson1988327,feingold,cartwright1,cartwright2,cartwright3}.
The resemblance of spiral and target patterns in crystal growth to such patterns is not mere accident.
Here we show with a coupled map lattice model that the classical BCF model of crystal growth is also an excitable medium.  As well as tangential growth of a solid crystal, we demonstrate that the excitable paradigm holds and the BCF model may be applied to layer by layer growth of a liquid crystal, which has recently been noted in the biomaterial nacre \cite{nacrepaper}.

\section{The BCF model}

Crystals grow either layer by layer, in tangential growth, or continuously, in normal growth 
\cite{Chernov:2003p2985,McPherson:2003p2958}.  In tangential growth the crystal surface is smooth on a molecular scale and consists of crystallographic planes of layers of growth units (atoms, ions, or molecules depending on the crystalline material). Burton, Cabrera, and Frank described how incomplete layers are bounded by steps, and incomplete rows of growth units along a step end at kinks. New growth units are incorporated at kinks. New steps are created only by  two-dimensional nucleation: the creation of a new island layer on top of an existing layer. The other mode of growth of such a layer structure is by a screw dislocation on the crystal face. In this case, a step propagates upwards in a helicoidal pattern, like a spiral staircase. A general point on the surface is then part of an already formed lower layer, and is quiescent, but receptive to the nucleation of a new layer above it given a large enough perturbation, that is to say a fluctuation that concentrates sufficient material nearby to nucleate a new surface. The edges of new layers are where new material is generally incorporated, while just after a new layer has formed at some point there is a refractory period during which another new layer cannot be formed, because any new material will of preference be incorporated at the growing edge nearby.

\section{Coupled map lattice model}

\begin{figure}[b]
  \includegraphics[width=\columnwidth]{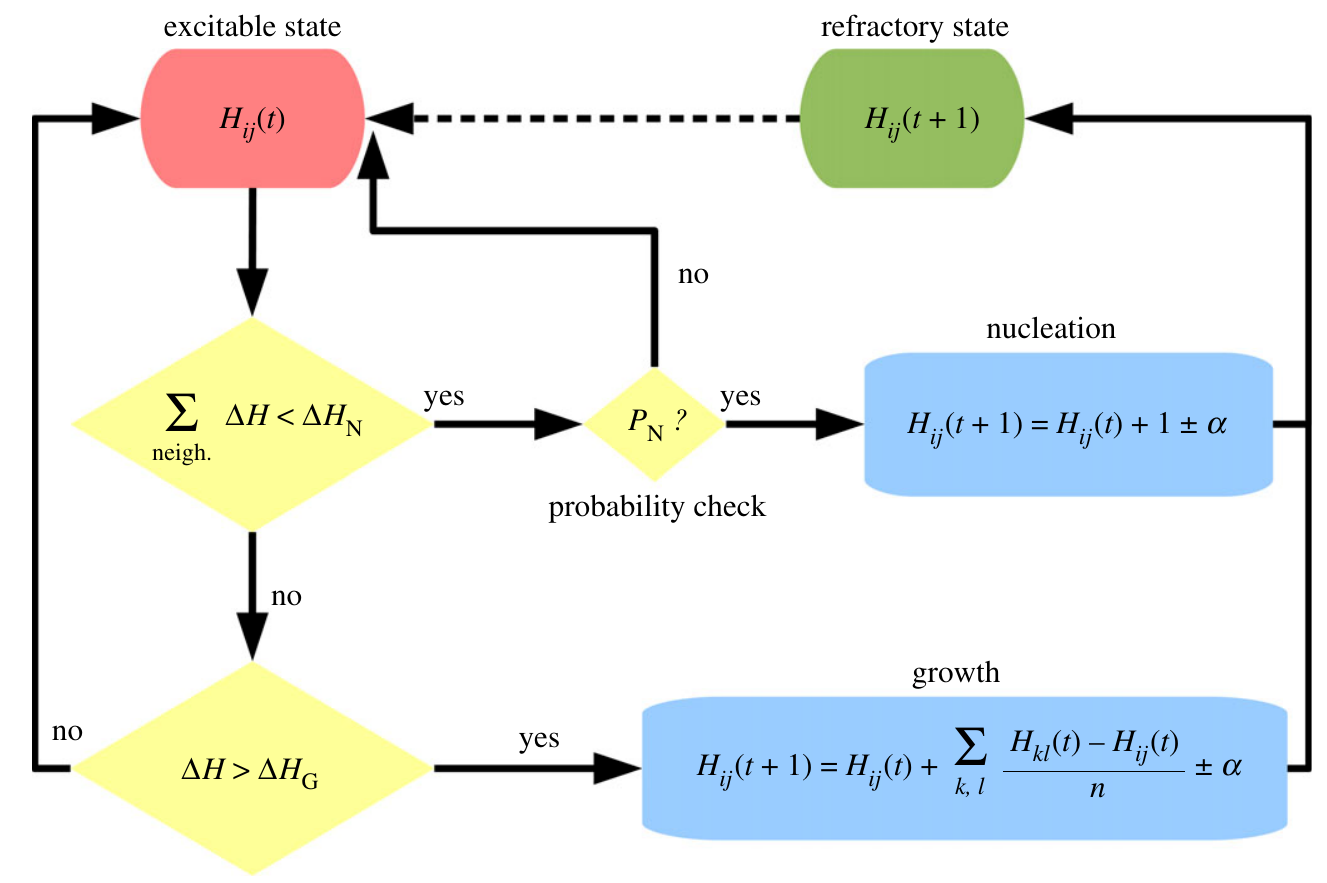}
  \caption{Flowchart of our coupled map lattice model combining the principles of tangential crystal growth and excitable media.}
  \label{flowchart2}
\end{figure}

Layer growth proceeds by the incorporation of individual growth units --- it is essentially a discrete process --- so we employ a discrete model.
(An early argument for introducing such discrete models into crystallography was given by Mackay \cite{mackay}.)
Thus we construct a minimal coupled map lattice model of the growth dynamics (Fig.~\ref{flowchart2}). The surface is divided up into cells, which may be thought of as the size of the growth units that compose the crystal. In order to avoid anisotropic growth, we use a randomized grid, as introduced by Markus and Hess \cite{Markus:1990}, and, defining a neighbourhood radius, $R$, the neighbours are all elements within this radius
$$
|r_{ij}-r_{kl}|< R,
$$
where $ij$ and $kl$ are the cell coordinates on the bidimensional lattice and we use the length of one cell as the length unit.
The parameter $R$ will influence the separation between terraces on the growing crystal surface, and so physically we can associate it with the surface diffusion of the depositing particles. 
Each cell has an associated height, $H_{ij}$, initially zero and updated at the end of each time iteration. $H_{ij}$ is a continuous variable into which a random component is introduced, as we show below, since there has to be certain variability for defects to occur.
Cells can be considered to be in an excitable state or in a refractory state. The excitable state means that the cell may nucleate a new island or add new material at a growth front. The condition for nucleation is that the cell in question must be on a flat surface, meaning that the height difference with its neighbouring cells must be smaller than a certain margin, $\Delta H_N$:
$$
\sum_\textrm{neighbors}\Delta H<\Delta H_N.
$$
This parameter will define the frequency of nucleations and target patterns. In crystallization it is related with the supersaturation or supercooling of the precipitating material.
The condition for growth is that the cell must be at the edge of a growth front, meaning it has at least one neighbour with a height difference larger than a certain threshold, $\Delta H_G$:
$$
\Delta H > \Delta H_G.
$$
This parameter is responsible for the appearance of screw dislocations. Physically there can be several causes for this kind of defects, such as the flexibility of the bonds of the lattice, the heterogeneity of growth units (for example in protein- and virus-crystal growth, and in liquid crystallization), and the presence of impurities.
If the cell does not fulfil either of these conditions, then it is considered to be in a refractory state. The duration of this state can be altered by increasing or decreasing the radius of the neighbourhood; a larger radius implies more neighbours and hence a longer refractory period, which is reflected in the separation between terraces in the spiral and target patterns.
However, a larger neighbour radius would also imply a higher probability of being at the edge of a step. This is reflected in the front thickness, meaning that the number of cells that experience growth in each iteration is greater, giving the appearance that the front is spreading faster. But since our time units (iterations) are completely arbitrary, the kinetics of crystal growth are beyond the focus of our model.

If the cell is in a position to nucleate, it must first pass a probability check with probability $P_N$, as nucleation is a stochastic process. $P_N$ is generally small ($\ll1$). If the check is positive and nucleation takes place, the height of the cell is increased by 1 plus or minus a small random factor $\alpha$:
$$ 
H_{ij}(t+1) = H_{ij}(t) + 1 \pm \alpha. 
$$
On the other hand, when growth takes place, the height of the cell is increased by the mean height difference with its higher neighbours:
$$
H_{ij}(t+1) = H_{ij}(t) + \sum_{k,l} \frac{H_{kl}(t)-H_{ij}(t)}{n} \pm \alpha,
$$
where $(k,l)$ are the coordinates for each higher neighbour and $n$ is the total number of higher neighbours, which depends on $R$. 
This growth algorithm is performed simultaneously for all cells, and the process is iterated.

\begin{table}[ht]\footnotesize
\caption{Model parameters} 
\centering 
\begin{tabular}{c p{4cm} p{3.1cm} p{3.3cm} } 
\hline\hline 
 & Description & Influence & Physical meaning \\ [0.5ex] 
\hline 
$\alpha$     & Stochastic term & Lattice defects & Impurities \\ 
$\Delta H_N$ & Height margin to consider flat enough to nucleate & Nucleations and target patterns & Supersaturation \\
$\Delta H_G$ & Height threshold to consider the base of a step & Screw dislocations and spiral patterns & Particle shape variability  \\
$R$          & Radius of the cell neighbourhood & Terrace separation & Surface diffusion \\ 
[1ex] 
\hline 
\end{tabular}
\label{table:parameters} 
\end{table}

\section{Results}

\begin{figure}[b]
  \includegraphics[width=\columnwidth]{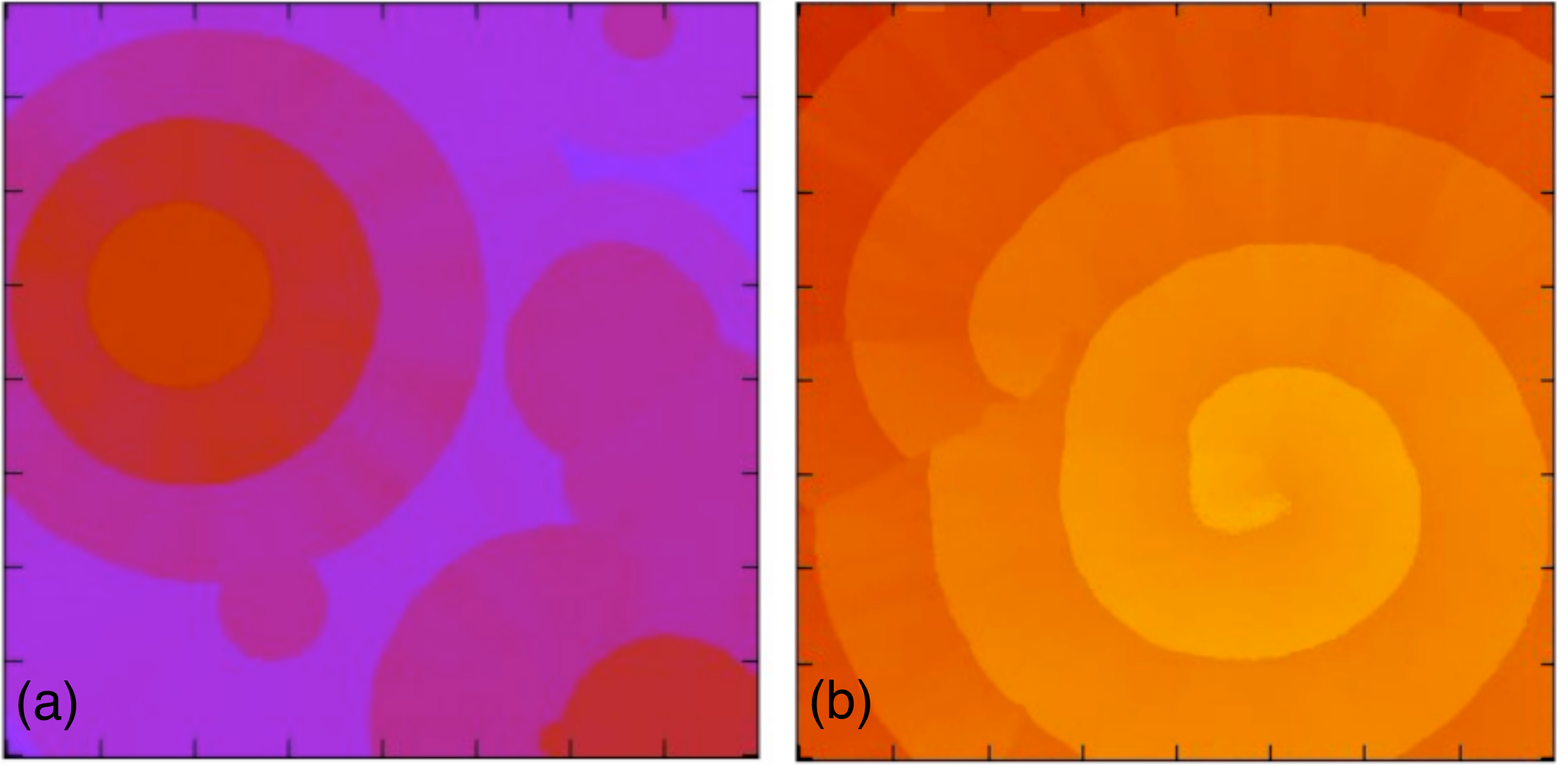}
  \caption{(a) Target pattern ($\alpha=0.08$, $\Delta H_N=0.1$, $P_N=10^{-3}$, $\Delta H_G=0.8$, $R=10$). (b) Spiral ($\alpha=0.05$, $\Delta H_N=0.06$, $P_N=1\times 10^{-6}$, $\Delta H_G=0.8$, $R=6$). The colour scheme represents height.}
  \label{patterns}
\end{figure}

When a nucleation event occurs it automatically inhibits further nucleation in its neighbourhood and all new material will be added at the edge of the newly formed island. If this island can grow through several time steps without greatly varying its height, a second island may nucleate on top of it. If this process continues periodically it will produce a target pattern (Fig.~\ref{patterns}a). In order to produce this pattern we set a broad nucleation margin $\Delta H_N$ to favour nucleation and a low stochastic term $\alpha$ to avoid rough surfaces. In crystallization these parameters would respectively translate to a high supersaturation or supercooling of the precipitating material and a low shape variability of its growth units. 
Although we can currently only describe the qualitative relation between the model and the physical parameters, understanding this behaviour is necessary in order to obtain a quantitative relation.

Another common pattern in both excitable media and crystal growth is the spiral (Fig.~\ref{patterns}b). When two growth fronts with slightly different heights collide, a portion of one may overlap the other, and continue its growth revolving around the dislocation centre. These patterns are produced with a small active neighbour threshold $\Delta H_G$, which allows growth fronts to overlap instead of annihilating each other, and a larger stochastic term $\alpha$ that introduces variability in the front height. 
Physically these parameters would imply a crystalline lattice with high tendency to create dislocations of whatever type, whether through the presence of impurities or due to the size and shape variability of the growth units.

\begin{figure*}[tb]
  \centering
  \includegraphics[width=0.8\textwidth]{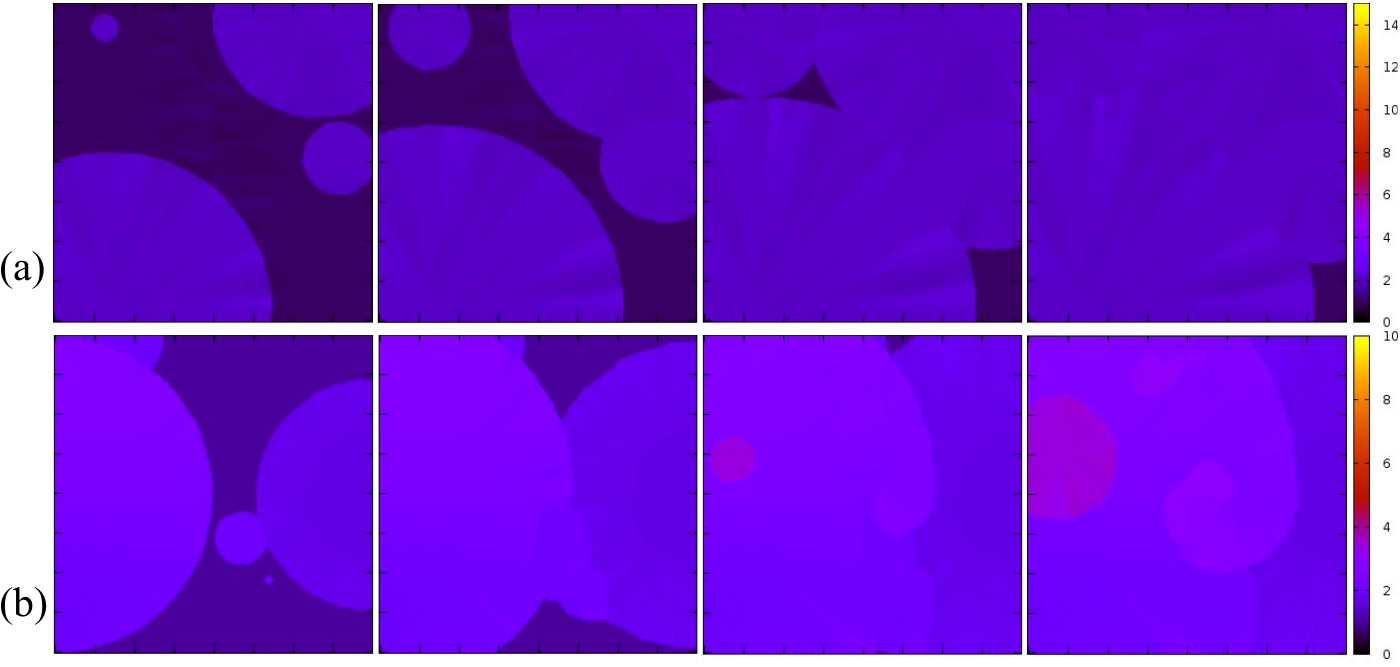}
  \caption{(a) Two growth fronts collide and annihilate each other ($\alpha=0.05$, $\Delta H_N=0.05$, $P_N=10^{-6}$, $\Delta H_G=0.8$, $R=6$).
	   (b) Partial overlap of two growth fronts produces a screw dislocation ($\alpha=0.05$, $\Delta H_N=0.05$, $P_N=10^{-6}$, $\Delta H_G=0.6$, $R=6$).}
  \label{sequence}
\end{figure*}

An important aspect of the behaviour of excitable media is that excitable waves cannot pass through each other, but rather destroy each other, as the refractory period does not permit wave propagation to continue; nor do waves reflect from boundaries. In the well-known excitable model of forest fires, for example, a second fire cannot pass until the vegetation burnt by the first has grown back, and excitable cardiac cells cannot fire again until they have recovered from their earlier firing. 
In Fig.~\ref{sequence}a we see a sequence of two growth fronts colliding and annihilating each other. In this case the height of both fronts is very similar, as the stochastic term $\alpha$ has been set to be small, and the threshold for overlapping, $\Delta H_G$, is too large for one front to grow over the other. With a higher variability in the front height and imposing a less restrictive growth threshold, one of the fronts may absorb the other, producing an edge dislocation, or it may partially overlap it, producing a screw dislocation (Fig.~\ref{sequence}b).

\begin{figure*}[tb]
  \includegraphics[width=\textwidth]{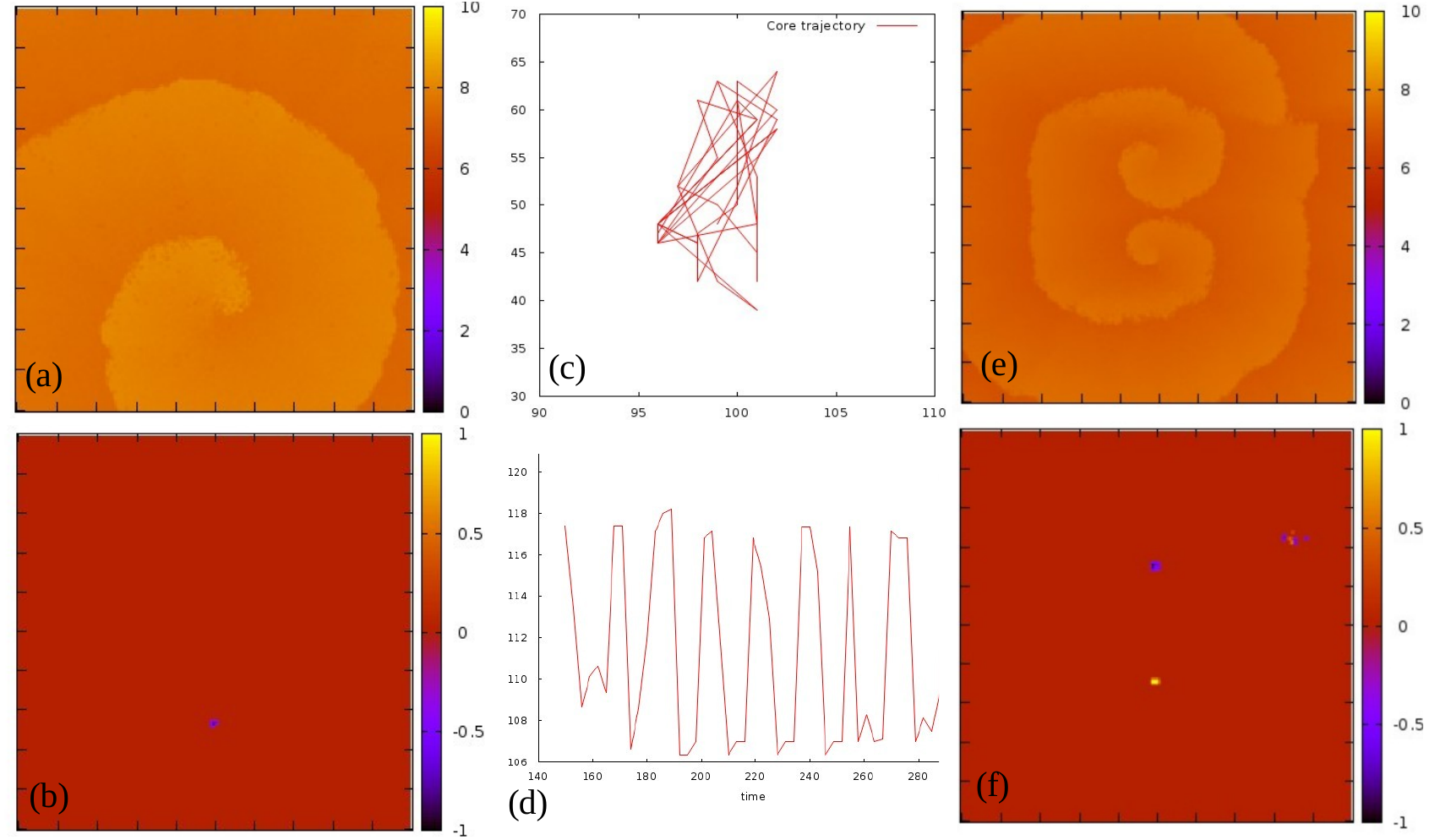}
  \caption{(a) Spiral created by a screw dislocation ($\alpha=0.05$, $\Delta H_N=0.05$, $P_N=10^{-7}$, $\Delta H_G=0.6$, $R=5$).
	   (b) Burgers vector calculated on the same map. The value of the vector is null everywhere except at the centre of the spiral.
	   (c) Trajectory of the centre of the spiral during 150 iterations.
	   (d) Distance from the origin to the centre of the spiral versus time during 150 iterations. 	   
	   (e) Double spiral pattern created by screw dislocations ($\alpha=0.05$, $\Delta H_N=0.05$, $P_N=10^{-6}$, $\Delta H_G=0.6$, $R=4$).
	   (f) Burgers vector calculated on the same map. The sign of the vector at the spiral cores is defined by the rotation sense.}
  \label{burgers}
\end{figure*}

An unusual aspect of our growth model compared to other excitable media is that here the third dimension of space --- outwards from the material --- is equivalent to time. Thus one can see in three dimensions the trajectory of the core of a spiral --- a screw dislocation in the material.
We can follow the position of this core by calculating the Burgers vector \cite{burgers}, the anholonomy in the lattice.
The traditional manner of calculating this vector is by drawing a Burgers circuit around the defect  and comparing it to the same circuit in a perfect lattice. But we cannot do this in our case because the lattice is not perfectly periodic: it has a random component on the plane and a continuous coordinate, $H$, in the third dimension.
Instead we take the following approach: We follow a square circuit around each cell looking for large height discontinuities ($>\Delta H_G$) that indicate the presence of a step. We accumulate the height differences around the whole circuit, adding when going up a step and subtracting when going down, and the resulting number is the magnitude of the Burgers vector in the vertical direction. 
If there is no defect around the considered cell, then the Burgers vector should be zero. But, if we are at the centre of a screw dislocation, the magnitude of the vector will be different from zero and its sign will mark the sense in which the spiral rotates (Figs~\ref{burgers}a and b). 
As the spiral continues growing, the position of its centre oscillates back and forth following a roughly straight trajectory (Fig.~\ref{burgers}c). 
When we plot the distance from the spiral core to the origin of our map in this case, we can see that the oscillation is periodic and we can estimate the amplitude ($\sim$12 cells) and frequency ($\sim$1 cycle per 20 time units) of the oscillation  (Fig.~\ref{burgers}d).
In Figs~\ref{burgers}(e, f) we can see another example where we use the Burgers circuit to find the two cores of a double spiral; a so-called Frank--Read source \cite{frank}.
The sign of the Burgers vector at the spiral cores is related to the rotation sense; a negative vector means clockwise rotation, and positive anticlockwise. 

\section{Discussion}

Tangential crystal growth involving spirals and target patterns has been intensively studied in the decades since Burton, Cabrera and Frank's pioneering work~\cite{Chernov2004499}. Falo et al.~\cite{falo} used a Langevin approach; a sine-Gordon equation plus noise to simulate spiral growth plus nucleation. Aranson et al.~\cite{Aranson:1998p2820} pursued a Ginzburg--Landau approach for spiral growth. And Smereka \cite{Smereka:2000p2800} took a spiral-growth continuum approach to BCF.  
More recent efforts include the application of phase field models that allow the quantitative calculation of physical properties such as kinetic coefficients~\cite{Uehara2003251} and the computation of more complicated morphologies such as dendrites~\cite{Chen2009702}.

\begin{figure*}[tb]
  \includegraphics[width=0.95\textwidth]{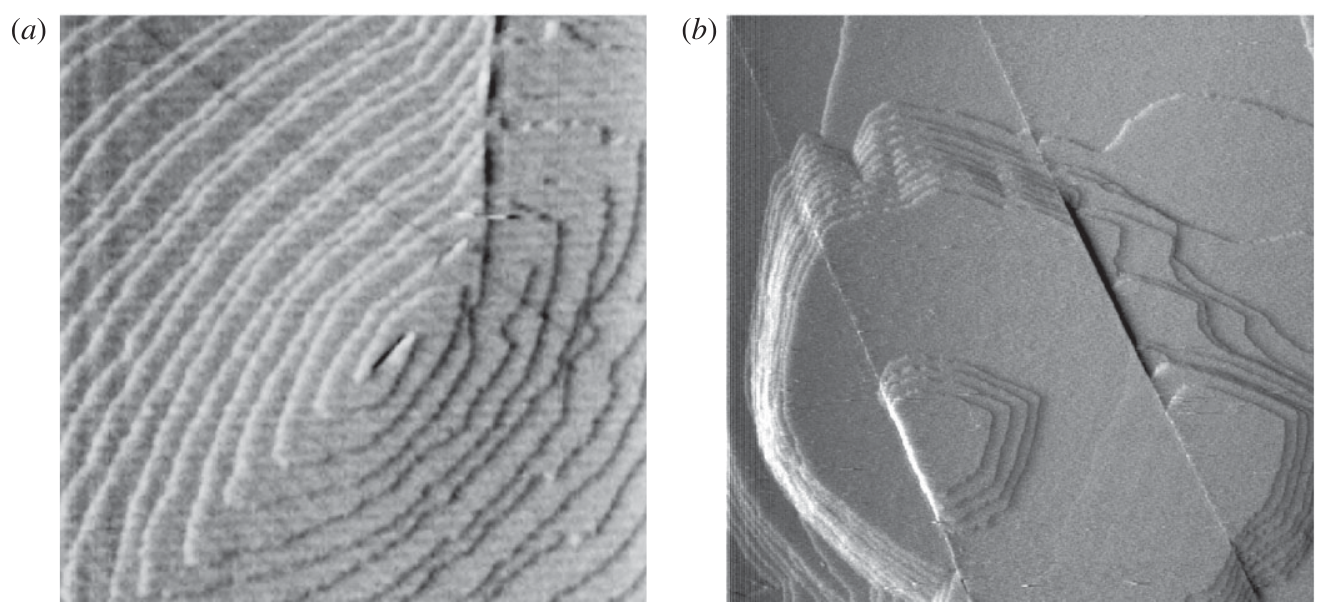}
  \caption{Spirals and target patterns in crystals: (a) spiral pattern on the surface of anglesite (PbSO$_4$) growing on a celestine (SrSO$_4$) crystal; scan area is $10\times 10$ $\mu$m ; (b) target pattern on the surface of a calcite (CaCO$_3$) crystal growing in the presence of magnesium;  scan area is $6\times 6$ $\mu$m. Both images were taken during growth with AFM in contact mode \cite{pina}.           
           }
  \label{afm}
\end{figure*}

In contrast with the above-cited works, our discrete model puts both growth modes, of spiral growth and island nucleation, on an equal footing. In other words, we use a coupled map lattice model instead of partial differential equations because we can produce both spiral and target patterns without requiring any other symmetry breaking elements.
The connection that we demonstrate here with the theory of excitable media has not previously been made. The connection is potentially of great utility to both fields, as excitable media have a series of theoretical models that can provide useful insight into crystal growth processes, while crystal growth is studied in much quantitative detail and can provide data with which to study it as an excitable medium. It is true that there are important crystal-growth phenomena that are not present in the minimal excitable-medium model; the Schwoebel effect leading to step bunching, etc \cite{pimpinelli,RevModPhys.82.981}. These act to make some crystal surfaces behave rather differently from the simple model. However many crystal surfaces do show the surface morphologies that reveal that this excitable dynamics is acting, as we may appreciate from scanning electron microscope and atomic force microscope images, such as those of macromolecular crystals \cite{Chernov:2003p2985,McPherson:2003p2958,pina}; see Fig.~\ref{afm}.

\begin{figure*}[tb]
  \includegraphics[width=0.95\textwidth]{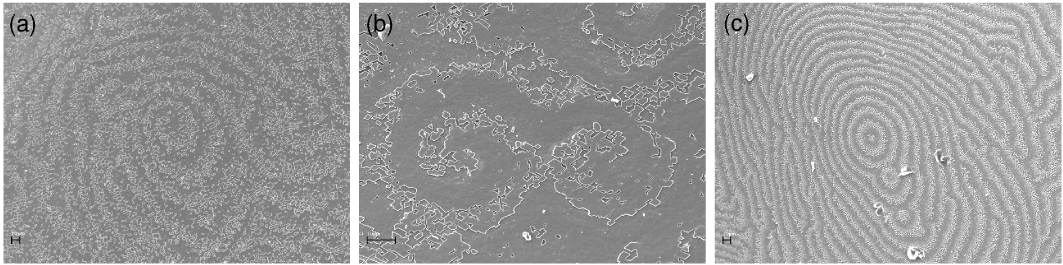}
  \caption{Spirals and target patterns in nacre; scanning electron micrographs of the growth surface of bivalve nacre: (a) Spiral (\emph{Isognomon legumen}) (b) Paired spirals (\emph{Pteria avicula}); (c) Target pattern (\emph{Pteria avicula}).}
  \label{nacre}
\end{figure*}

The possibility that liquid crystals might grow layer by layer in the manner of a solid crystal seems to have been unexploited by man; not however by nature. In fact our interest in an excitable-medium description of layer growth arose out of our work with the biomaterial nacre \cite{interface}. Nacre, or mother of pearl, is the iridescent material that makes up the inner layer of many mollusc shells, and also the pearls that form around a foreign body within the mollusc. It is a material that has been much studied, because of its beauty, its strength, and its interest for biomimetics as a complex material self assembled outside the cells of the organism that produces it. We have recently shown that nacre construction begins with the self assembly into a cholesteric liquid crystal of rod-shaped crystallites of the polysaccharide $\beta$-chitin \cite{nacrepaper}. Liquid crystals are usually formed in laboratory experiments by altering a global experimental variable, be it temperature, concentration, electric field, etc, that causes the whole of the experimental domain to crystallize at once. But the mollusc uses another mode of construction, in which the liquid crystal is built up layer by layer, in the manner of a solid crystal.  It is for this reason that we find in nacre (see Fig.~\ref{nacre}) the characteristic spiral, target, and labyrinthine patterns we have discussed above in the context of solid crystals. It is our opinion that understanding this equivalence between layer growth in solid and liquid crystals --- and it appears probable to us that nature uses layer growth of liquid crystals ubiquitously --- might provide useful new ideas for technological applications of liquid crystals.

\vspace{0.5cm}
We would like to thank Carlos Pina for his permission to use the AFM images of Figure 5.
We acknowledge MINCINN (Spain) projects FIS2010-22322-528C02-02 and 
CGL2010-20748-C02-01.


\end{document}